# Evolving Cognitive Architectures

Alexander Serov

Research Group "Automatic Intelligent Data Acquisition" (RG AIDA), Russian Federation

aidaresearchgroup@gmail.com

**Abstract**

This article proposes a research and development direction that would lead to the creation of next-generation intelligent technical systems. A distinctive feature of these systems is their ability to undergo evolutionary change. Cognitive architectures are now one of the most promising ways to create Artificial General Intelligence systems. One of the main problems of modern cognitive architectures is an excessively schematic approach to modeling the processes of cognitive activity. It does not allow the creation of a universal architecture that would be capable of reproducing higher nervous functions without using a predetermined set of perception patterns. Our paper proposes an evolutionary approach to creating a cognitive architecture. The basis of this approach is the use of a functional core, which consistently generates the intellectual functions of an autonomous agent. We are considering a cognitive architecture that includes components, the interaction of which ensures the evolution of the agent. The discussion of the development of intelligence is carried out using the conceptual apparatus of semiotics. This allows us to consider the task of developing cognitive functions as a problem of establishing a connection between the Merkwelt and the Werkwelt through the creation of the Innenwelt. The problem of early postnatal ontogenesis is investigated on the basis of the theory of constructivism: we discuss the requirements for the functional core and its composition, as well as the mechanism that initiates the process of cognition.

## 1. INTRODUCTION

Humanity's technological development can be seen as strengthening the resilience of human society in the outside world. A significant number of technologies are being developed to protect us from the detrimental effects of changing natural conditions: extremely low or high ambient temperatures, floods, sudden temperature changes, prolonged periods without rain, earthquakes, lightning strikes, etc. The technical means embodying these technologies are currently being actively supplemented by various "smart" sensor systems, AI agents, and robotic systems, making our environment as comfortable as possible. All these technological means can be called second nature. They are designed to create a comfortable living environment for modern humans.

Second nature devices are currently evolving toward the increasing use of advances in artificial intelligence theory. This allows us to speak of the increasing adaptability of technologies both to everyday human activities and to the conditions of their existence in the outside world. One of the most important areas of development of second nature means is the monitoring and characterization of both local and global natural processes. The ability to predict changes in the human environment is necessary for the timely technological compensation of the part of these changes, which negatively impact human society. Currently, this area of scientific and technological development could be compared to the creation of specialized systems in which research tasks would be divided between humans and a computer system whose architecture is based on AI principles. Clearly, the nature of the division of these functions is determined not only by the purpose of such systems, but also by the degree of intelligence of the computer system.

Today, it can already be concluded that Large Language Model (LLM) technologies make it possible to automate many tasks formulated in the fields of development and maintenance. The success of using pre-trained text models is due to a significant reduction in the cost of solving practical problems associated with rapid searches and the creation of summaries of search results. LLMs have become widespread due to the ability to interact with them in natural language. Attempts to use computational technologies based on LLM can be found today in areas of analytical activity that are quite distant from textual data in terms of information representation (e.g., [1]). We believe these attempts are unpromising. There are several reasons for this. We recognize that the use of numerical methods derived from LLM methodology is a consequence of the insufficient development of other methods for automating analytical activity. The use of pre-training methodology in the field of Deep Learning encounters a number of fundamental difficulties. Language (whether artificial or natural) is based on a discrete representation of information. At the same time, most areas that require the automation of analytical activity must process measurement results with a continuous value scale. Language is a homogeneous representation of information. At the same time, there is a practical need to work with multiple data streams, the structuring of information in each of which may be unique. Pre-training models allows us to work with static data. However, in the real world, we must solve problems characterized by constantly changing contexts and significant dynamics of the surrounding world. Methodologies based on pre-training are always based on a basic set of elements representing knowledge. These methodologies automate the processing of data about the world as it is perceived by humans. At the same time, AI technologies are capable of, and should evolve toward, creating fully autonomous technical systems capable of self-evolving. This would expand the range of our perception of the world beyond what we currently possess. This manuscript is devoted precisely to this problem - the creation of autonomous intelligent agents capable of self-evolving, accumulating knowledge about the world around them and modifying themselves.

## 2. COGNITIVE ARCHITECTURES

Currently we are at the stage of active implementation of cyber-physical systems in almost all areas of modern production. The development of smart technologies is determined by an increase in the automation of devices, a change in the structure and nature of interaction between them, and the active use of self-monitoring tools. Industry 4.0 involves the use of intelligent agents integrated into a single information space and capable of interacting with each other without human intervention. The creation of such a system imposes a number of requirements on the characteristics of the devices. One of the most significant is the ability for autonomous intelligent data processing. Other requirements include minimizing human involvement in the device reconfiguration process. The creation of self-configurable, self-healing devices able to perceive the environment and optimize their own performance in purposeful activities is a problem in the field of Cognitive Science.

The term Cognitive Architecture was first used in the second half of the twentieth century [2, 3]. It was defined as a set of basic principles of operations of a cognitive system. This term is used now both in theoretical research, and in technological projects devoted to the development of Artificial Intelligence (AI) systems. Currently, there are known more than 300 cognitive architectures [4, 5], which solve the problems of modeling the processes of cognitive activity in various ways.

Two main directions of the development have emerged during more than 40 years of the existence of research in the field of cognitive architectures [6]. The first of them is associated with the creation of systems that implement the ideas of Weak AI and intended for use in a narrow field of applications. In these systems only certain aspects of cognitive activity are reproduced. For example, in the Conscious-Emotional-Learning Tutoring System [7], three learning mechanisms are integrated, the main of which is associated with emotions. Another example is the For the Right Reasons project [8], which is dedicated to the creation of an architecture that simulates the decision-making process based on the use of an expert knowledge.

The second direction of a development in the field of cognitive architectures implements the ideas of Artificial General Intelligence (AGI). The main goals of projects related to this area are the development of a theoretical foundation for the creation of highly intelligent systems and the development of architectures that most completely reproduce cognitive functions. Each project belonging this direction has its own contribution to a practice of intelligent systems creation. The difference in theoretical positions used as a basis in each project development may be explained by the lack of a generally accepted theory that would describe cognitive activity. This difference leads to different implementations of intelligent functions. For example, the project dedicated to the development of the Learning Intelligent Distribution Agent (LIDA) [9] is based on the implementation of the Global Workspace Theory [10]. The Connectionist Learning with Adaptive Rule Induction On-line project [11] is based on the integration of models developed based on the results of research in the field of psychology. The project related to the development of the Adaptive Control of Thought - Rational (ACT-R) architecture [12] was developed on the basis of the Rational Analysis methodology.

The creation of a unified theoretical foundation for modeling cognitive processes would make it possible to concentrate a research in a single subject area. One of the attempts to create such a foundation was undertaken in [13] based on the synthesis of the cognitive architectures ACT-R, Sigma, and Soar. The authors of [14] proposed the methodology used in the LIDA project [10] as an alternative way for creating AGI systems. The problem of unified theoretical foundation is currently one of the most urgent: its solution would significantly reduce the time necessary for the transition to Industry 4.0. However, this is not the only problem that hinders the development of intelligent technologies. A promising goal in the field of AI is the development of devices that would not be inferior to a human in their intellectual performance [15]. In the next sections of this paper, we will touch upon issues related to achieving this goal.

## 3. AUTONOMOUS COGNITION

It is currently envisioned that the technologies of the future will be based on a distributed system of intelligent devices. Thereby a special place in a research on cognitive technologies is occupied by the topics related to the creation of autonomous systems.

3.1 Simulating intellectual activity

One of the most promising areas for the development of autonomous agents is currently the direction of Biologically Inspired Cognitive Architectures (BICA) [5, 16]. BICA approach is based on the idea to emulate the type of data processing which is typical for living beings [17].

Cognitive architectures that are currently being created can be divided into two main types, which differ in their approach to modeling intellectual activity: Reactive and Deliberative architectures. Deliberative architectures implement Good Old-Fashioned Artificial Intelligence methodology [19, 19]. The workflow on the processing of data by these architectures can be described as a cycle that includes the stages of sensing, modeling, planning and acting. Most of modern projects devoted to the development of autonomous cognitive systems are based on this type of architecture.

An approach to the development of autonomous agents based on Reactive architecture was formulated in [20]. It is based on the use of a certain set of behavior models. An agent reacts to a change of its existence in the world by means of these models. As an example, an architecture based on the use of a hierarchy of state machines was presented. Each of these state machines solves a separate control problem. The levels of control in the system are linked by a subordination scheme, the mechanisms of which include suppression and inhibition. The title of the article [20] provides a key to understanding the approach used at creating the architecture of the control system. The work criticizes the Deliberative architectures methodology: during development, several subsystems are artificially singled out with the subsequent determination of the interface of interaction between them. For example, agents described by architecture [21] contain the following modules: Perceptual Buffer, Short-term Belief Memory, Short-term Goal Memory, Motor Buffer, Long-term Skill Memory, Long-term Goal Memory, Long-term Conceptual Memory. This entails the need to introduce an information presentation layer that must be used to provide said interaction. As a result, this leads to the imposition of a number of restrictions on the developed system, which in the general case have no relation to the simulated reality.

3.2 Simulating data processing

Historically, there have emerged three main approaches which describe processing of data by agents in different ways: top-down (symbolic), bottom-up (emergentist, subsymbolic), and hybrid approaches [22, 23]. The work of symbolic architectures is based on the concept that processing of data by cognitive systems occurs through the manipulation of information represented in a symbolic form.

Each symbol has some meaning, since it reflects a certain aspect of reality. The logic of data processing consists in the recursive decomposition of the problem being solved into a set of simpler ones. Cognitive architectures based on this approach include, for example, EPIC [24], ICARUS [21], COGNET [25]. This line of research and development has the richest history. Its obvious advantage is the ability to use the formal logic in modeling the processes of knowledge extraction and information processing.

The subsymbolic approach is based on the idea that the symbolic representation of information arises from the processing of signals received by the agent from its sensors. Cognitive architectures that are built on the basis of this approach among others include HTM [26], DAC [27], BBD [28]. Most often, when implementing agents based on the subsymbolic approach, neural network models are used. The strengths of this type of architecture are manifested in the processing of multidimensional datasets through the ability of recognizing patterns. These architectures also are characterized by easily implemented machine learning methods, in particular, Reinforcement machine learning schemes.

The disadvantages and advantages of the symbolic and subsymbolic architectures are opposite in some sense. By means of the symbolic architectures, it is easy to implement the required data processing logic. However, this logic is based on a predefined symbol set. The subsymbolic architectures can be used to create symbolic representations based on sensory data processing. However, with their use it is difficult to implement high-level mental functions (abstract reasoning etc.). Attempts to combine the advantages of the mentioned architectures and to avoid their shortcomings led to the creation of hybrid architectures. Hybrid cognitive architectures are based on the integration of bottom-up and top-down approaches.

According this, two information processing subsystems are involved in the integration: subsymbolic and symbolic ones. The most famous hybrid architectures are LIDA [9, 10, 14], CLARION [11, 29], SOAR [30], ACT-R [12]. It is currently assumed [31] that AI systems having a level of intelligence comparable to that of a human will occur on the basis of the combination of symbolic and subsymbolic data processing methods.

3.3 Hard problems

Cognitive activity is determined by the ability to extract new information. In this case, an important role is played by the basis on which this information is extracted. Early advances in symbolic architectures were associated with the use of symbols from a predefined alphabet. In accordance with this, the decomposition of sensory data by the agent during operations of analysis cannot go beyond the elementary set of characters. Similarly, information synthesis operations should be based on the use of a predefined alphabet. Specifying the set of elements (into which the sensory data stream is to be divided) implicitly configures the autonomous agent and therefore bounds its cognitive capabilities.

Data processing by subsymbolic architectures is not characterized by pre-defined elementary symbols. However, the use of these architectures is associated with an implicit predetermination of the meaning that sensory data have. An example is the method used to train neural networks to recognize the image of a specific object. A set of photographs containing this object is used as the initial data. The use of the unsupervised machine learning method makes it possible to extract a set of characteristic features, the use of which will further enable the identification of the object. Thus, in this case, the information content of the numerical model is predetermined.

The use of both symbolic and subsymbolic architectures, which are characterized by these features, limits the degree of agent autonomy and its cognitive capabilities. The advantages of both architectures can be combined by finding a method by which the autonomous agent would be able to independently define the alphabet based on the perception data. The transition from the subsymbolic to the symbolic level of representation can be characterized as extracting information from sensory data, structuring the data of perception. It must be characterized by such a transformation, the result of which has a meaning for the autonomous agent. The problem of finding such a data transformation has been called the Symbol Grounding Problem (SGP) in the literature [32].

SGP is a problem of semantics that data must acquire as a result of processing by a cognitive architecture. Its solution can be represented as "an unsupervised process of establishing a mapping from huge, noisy, continuous, unstructured inputs to a set of compact, discrete, identifiable (structured) entities, i.e., symbols" [33]. Despite the fact that during the existence of the problem several solutions have been proposed (see, for example, [27, 28, 34–36]), a general solution has not yet been found [37, 38]. This is usually associated with the implicit use of pre-installed semantic resources into the cognitive system.

The importance of the SGP solution at the creation of new generation cognitive architectures can be explained as follows. Modern cognitive systems operate on the basis of a syntactic manipulation of symbols. However, this manipulation is not enough to generate a meaning. Intelligent data processing is determined by the process of thinking, which manipulates meanings by referring to their symbolic representation. Thus, we can say that SGP generates the following interrelated problems: how to generate a symbolic representation; how to link a meaning and its symbolic representation; how to process meanings [39].

We can conclude that in the cognitive systems of the future, the main processed element should be not the symbol, but the meaning. This gives rise to another problem [37, 39]. It is associated with the requirement that when solving SGP, neither external nor internal pre-defined semantic resources of an autonomous agent should be used. This requirement is fundamental at creating AGI agents. The main question at creating true cognitive agents is the way of data processing, during which the meaning is generated. The Symbol Grounding Problem, understood as the problem of the dynamics of processes characterizing cognition (especially) at the early stage of ontogenesis, is formulated as the Symbol Emergence Problem (SEP) [40, 41].

A common drawback of modern cognitive AGI architectures is the lack of implementation of the mechanisms for generating semantics in data processing. We consider this paper as a step in solving this problem. It is devoted to the problem of creating a cognitive architecture which is characterized by the absence of predetermined information about reality. A distinctive feature of the architecture is the ability to simulate the dynamics of cognitive abilities and characteristics of a cognitive agent at the first stage of postnatal ontogenesis.

In connection with the problem of semantic generation, it is necessary to recall the work [42], which proposed artificial network architecture called Transformer (the first sequence transduction model based entirely on attention). Unfortunately, models of this kind cannot be used in situations where a basic set of symbols is absent. The process of meaning formation is multistage and we will briefly discuss it below.

The content of this article is structured as follows. Section 4 is devoted to the introduction and discussion of a terminology. We use the terminology typical for semiotics research. This is in line with our goals for the following reasons. Most of the work on the subject of cognitive architectures is devoted to modeling cognitive functions. Their authors use modern knowledge about the psychology and neurophysiology of the brain and introduce structural elements of the architecture (i.e. intentional module, episodic buffer, etc.) on this basis. The terminology of semiotics allows highlighting a number of philosophical problems, the consideration of which is essential in matters of cognition. The use of semiotic terms together with the terminology prevailing in the field of Artificial Intelligence also makes it possible to consider the problem of cognition at the required level of abstraction. Section 5 is devoted to the use of the constructivism approach in the field of Artificial Intelligence. Section 6 includes the description of the general architecture of the cognitive agent control system. In this part of the paper, we discuss the problem of the agent's cognitive evolution based on the development of the functional core. The problem of early postnatal ontogenesis is investigated by us on the basis of the theory of constructivism: we discuss the requirements and composition of the functional core. Section 7 examines the driving force behind the cognitive development of autonomous agents. Section 8 concerns the simulation of the initial stage of mental development of the agents. Section 9 contains a discussion of the main results of our work.

### 4. SEP, PSYCHOLOGY AND SEMIOTICS

The Symbol Emergence Problem solution involves uncovering the way of creating a system of symbols, the use of which would allow an autonomous agent to perform its functions. The formation of a system of symbols allows optimizing the performance of these functions: it minimizes time and resource costs and maximizes the quality of performance. According to the analysis given in [37], SEP can be considered only if the agent uses internal resources that do not have a semantic component.

The formulation of SEP requires the specification of the initial mental state of the cognitive agent. Using the constructivist approach, we will assume that the agent does not have any knowledge about the world of embodiment. As we have already noted, this condition is necessary when considering AGI agents. The absence of knowledge about the world of the autonomous agent embodiment corresponds to the modern views of Developmental Psychology.

The need to simulate higher nervous functions requires an approach based on the results of research in the field of psychology. From our point of view, modern cognitive architectures are characterized by excessive reductionism: they excessively concentrate on the embodiment of knowledge about the neurophysiology of the brain (i.e. procedural memory, sensory-motor memory, episodic memory, semantic memory, etc.) and underestimate the role of dynamic factors in the formation of cognitive functions.

The creation of new generation cognitive architectures requires simulating the dynamics of the development of cognitive functions. The methods for processing sensory data are actively formed at the stage of sensorimotor development, which is the first stage of ontogenesis. According to the theory of the development of cognitive functions [43, 44], the child's thinking is characterized as peculiar, qualitatively different in its properties from the thinking of an adult. In particular, at the initial stage of postnatal ontogenesis, a child is characterized by solipsism and inseparability of the world and his/her own self.

Simulating mental phenomena requires the use of an adequate descriptive means. The main requirements for the means of description include the ability to consider mental phenomena in dynamics from the very beginning of sensory data processing; the presence of a universal terminology introduced to explain the phenomena of the subjective world; the integrated nature of the description of cognitive processes. The results of research in the field of semiotics can help to find such descriptive means. In investigations of the founders of semiotics there was introduced a system of concepts that can be used to describe the forming of signs and sign processes in living systems [45].

An autonomous agent exists in the real world. The world of the agent's embodiment, described by the term Umwelt (translated from German as environment) [46], is different from the real world. Umwelt is the self-centered world of an organism, the world as known, or modeled [45]. Umwelt is a subjective "slice" of the real world belonging to a representative of a particular biological species. The Umwelt of a living being consists of signs, data about which it receives from receptors and which it interprets based on experience, knowledge and the context of the tasks being solved. Alive being has the set of sensors that characterize the state of the Umwelt and the set of effectors that are used to make the change in Umwelt.

Umwelt is always unique because it characterizes a certain living being. But we can talk about the degree of Umwelten (plural from Umwelt) similarity, since the structure of the organism of living beings of the same species is of the same type. The Umwelt contains two parts, allocated from the surrounding reality by the sense organs and the organs of action: the perceptual world (Merkwelt) and the operational world (Werkwelt). Merkwelt and Werkwelt interact through the functional cycle (semiosis). Semiosis is defined as a "process in which something is a sign to some organism" [45]. The cognitive process is characterized by the creation of Innenwelt. While Umwelt is a personified part of objective reality, Innenwelt is "the world as represented in the sign system of an organism" [45].

Cognition of the world by a living being according Peirce's theory occurs due to a semiosis, the model of which is represented by the interaction of three components: object, representamen (sign) and interpretant [47]. In this triadic process, the representamen characterizes the object of the world in some way, and it is the correlant between the object and the interpretant. The interpretant in this relation is a sign that is generated in the interpreter (i.e. in the cognitive system) under the action of a representamen in order to interpret the object. It should be noted that the representamen and the interpretant cannot be identical, since the first belongs to Umwelt, and the second to Innenwelt. The presented process of semiosis describes only the first stage of interpretation.

The subsequent stages involve other representamens and interpretants, each of which belongs to the Innenwelt: the interpretant from the semiosis described above turns into a representamen in the subsequent stage of the interpretation process. Thus, the whole process of interpretation is expressed by a chain of propagating semiosis processes.

The above considerations make it possible to introduce the concept of meaning of the sign. The meaning of the sign is a response from the interpreter to the influence of the object that is represented by this sign, i.e. it is an interpretant corresponding to this sign. Thus, the meaning of the sign should be understood as the entire chain of semiosis processes that is generated in the interpreter under the action of representamen.

The model of semiosis represented above is typical for an adult living being. It is based on the fact that the Innenwelt is workable, i.e. on the ability of a living being to interpret the data of perception. The early phase of postnatal ontogenesis is characterized by the opposite situation. In this regard, a number of questions arise related to the solution of the Symbol Emergence Problem. One of them is the question about the origin of the first signs (symbols).

The results of research in the field of psychology [48] and neurophysiology [49] show that the decisive role in the intellectual development of living beings is played by the goals that must be achieved during activities. The formation of the Innenwelt is similar to the creation of a cognitive map that contains a symbolic representation of reality. The emergence of symbols and connections between them occurs due to the activity of a cognitive agent, i.e. due to the joint processing of data from the Merkwelt and Werkwelt domains. In physiological interpretation the emergence of a symbol corresponds to the process of neurogenesis, and it is always associated with the acquisition of meaning. The methodological apparatus that is most adequate in relation to the tasks of cognitive evolution is currently being actively developed in the field of biosemiotics [50].

## 5. CONSTRUCTIVIST APPROACH

Research in the field of Artificial Intelligence that investigates cognitive functions in their dynamics belongs to the field of Constructivist AI. These investigations traditionally are based on the Constructivist Psychological Theory proposed by Piaget. He created psychological constructs, on the basis of which ones it becomes possible to simulate cognitive functions. One of the central concepts used in Constructivist AI is Schema Mechanism [51]. Schema is defined as a triple $\{C, A, R\}$, where C is the context representing the situation in which the schema is applied, A is the action that the agent performs in the environment, R is the expected result of the action.

The main aspect of intelligence is adaptation, the goal of which is to achieve balance with the environment (homeostasis). Adaptation is an external expression of the internal processes inherent in cognitive agents. The role of these processes consists in unstoppable transformation of cognitive structures, which are described by schemes.

According to [52], the main components of adaptation are the processes of assimilation and accommodation. In the ideal case these processes balance each other. Assimilation and accommodation are the processes that transform the Innenwelt through its interaction with Merkwelt and Werkwelt. Assimilation is expressed through the incorporation of new components of the Merkwelt and Werkwelt into the work of the schemes already existing in the Innenwelt. Accommodation consists in modifying schemes in order to adapt them to existing conditions.

The indisputable advantage of the architectures developed using the Constructivist AI approach is their "innate" applicability in the dynamic modeling of intelligence development. The limitations in the applicability of this approach are related to the peculiarities of the interpretation of the underlying theory. Schema Mechanism [53] is easily implemented only in the case of symbolic architectures. In this case, the values of C, A and R must be predefined for the initial set of schemes. The disadvantages of this approach were discussed in Sect. 3.3. When creating hybrid architectures, the SEP solution method must be embedded in the sensory data processing path. This requirement follows from the need to define primary schemes.

Constructivist AI is currently using the operational concept of intelligence. The logic realized in this case obeys biological laws. As a result, the complex nature of the formation of cognitive processes is not taken into account: in addition to sensory and motor components it must include emotional and volitional components. Used approach entails a simplification or even the absence of a formulation of the task of achieving homeostasis when simulating the dynamics of intelligence development. All this taken together leads to limitations in the applicability of the developed methods during modeling high-mental functions.

## 6. ARCHITECTURE OF THE CONTROL SYSTEM

We assume that the functions of the control system can be split into two levels. First of them is directly related to the Merkwelt and Werkwelt. This is Umwelt interface layer. Data processing realized in this level depends on the embodiment of the agent. This level of architecture implements the functions of perception and direct control of the body. The second level of architecture is not directly related to agent embodiment. Methods of data processing at this level are abstracted and are intended for the implementation of higher mental functions.

### 6.1 Core components of the architecture

The architecture of the control system consists of the following core components: Perceptual component, Motor component, Intelligent component, Emotional component, Volitional component.

Each core component includes a functional kernel and elements that provide control. The kernel is responsible for the following set of functions:

- Generating control signals and transferring them to the elements of this component
- Providing the evolution of core component by creating new elements and changing the connections between existing elements
- Interacting with other core components of a control system

We assume that the elements of each core component are programmable devices with memory. Each element is able to process input data according uploaded program, generate an output signal based on the results of processing and store information about its state.

Perception, the ability to perceive, is the basis for the emergence of intelligence. Perceptual component provides the cognitive agent with data about Merkwelt. The process of cognition is impossible without embodiment and the ability to act in the world. Motor component is responsible for the transferring of information between core components of the architecture, and the low-level control of the agent's body. By the term "low-level control" we mean this control is not directly connected with mental processes. Low-level control includes the following functions: (1) automatic control of the processes responsible for the existence of the agent's body; (2) the control of effectors that are used for manipulations in Werkwelt.

Perceptual and Motor components are those architectural elements that provide the existence of the agent and its interaction with reality. These two architecture components implement the Umwelt interface layer of the agent control system. The process of cognition runs via the experience of Merkwelt perception and the experience of actions in Werkwelt. It leads to the creation of knowledge and its consolidation in certain patterns of behavior.

Intelligent component is responsible for creating, modifying, storing and managing mental models (symbols) that describe reality. Intelligent component implements Innenwelt.

According to [48], the emotions of a natural agent are integral characteristics. Each emotion is a reflection of the quality and magnitude of the need experienced by the agent. Every emotion also depends on the possibility of need satisfaction, which is assessed by the agent on the basis of individual experience. Emotional component is responsible for two interconnected aspects of data/information processing: agent learning, and patterns recognition. Autonomous agent learns in a self-supervised manner, and emotional component is a key element in this process. At the end of the natal ontogenesis, natural autonomous agent has an unformed Innenwelt. Emotional component participates in creating and categorizing hierarchically organized set of mental models. The process of Innenwelt creation includes programming of behavioral reactions. This programming is based on the categorization of high-level models. Emotional component is responsible for the creation of categories. It is also responsible for assigning labels of categories to mental models activated during processing of sensory data.

Cognition is the result of activities that have specific goals. Goal-setting, actualization of behavior patterns, control of the results of actions are carried out due to the architecture component that performs the function of general management. Volitional component coordinates the work of other core components of architecture. It is responsible for allocation of energy resources. One of most discussed problems in papers devoted to the philosophical problems of Artificial Intelligence is the issue of consciousness/self-awareness [54, 55]. In a more general formulation this issue is reduced to the so-called problem of Self [56]. Volitional component in the proposed architecture can be considered as a platform for experimental research in this area.

6.2 Interaction of core components of the architecture

Our approach to modeling cognitive activity can be illustrated by the interaction scheme presented in Fig. 1. The motor component is the central element of the component interaction scheme. This may be explained by the fact that information is transferred between core components by means of this component.

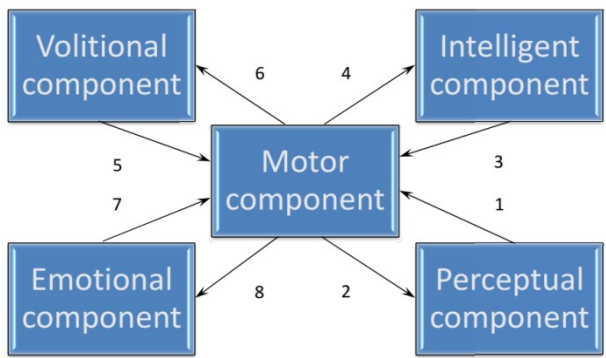

Fig. 1 Data/information streams between core components

Perceptual and motor components are directly integrated in the agent's body to provide control and monitoring of its state. Processes 1, 3, 5, 7 implement the transmission of data, information, or control signals from the definite core component. Processes 2, 4, 6, 8 are associated with transferring these to core components.

Processes 1→4 provide the transfer of sensory data to the intelligent component by means of the motor component. Notation a→b means that process a is followed by process b. These data are used by the agent for the learning and update of the mental state. Sending commands by the volitional component for the high-level control of the agent's body is implemented by processes 5→2. Processes 3→6 and 5→4 provide information exchange between the volitional and intelligent components. These processes are interconnected, interdependent. Intelligent component sends the results of data processing to the volitional component. In general case processes 3→6 can be associated both with processing of data received from the perceptual component (updating the state of Merkwelt), and with emulation of the thinking process (knowledge inference via Innenwelt). The purpose of processes 5→4 in the operation of the control system is twofold. Volitional component is responsible for the control of high-mental functions. So this set of processes is used for the control of information extraction or for the modification of Innenwelt. Processes 7 and 8 describe information flows that link the emotional component to other components of the control system. The emotional component is involved in all the processes characteristic to the cognitive activity. Processes 3→8, 1→8, 5→8 are responsible for the transmission of the information that is required to compute the value of the emotion. Processes 7→6 describe the transfer of the calculated value to the volitional component.

The emotional component is involved in the processes of higher nervous activity, performing four types of functions [48]. The reinforcing function of emotions is manifested in the process of the formation of conditioned reflexes (in particular, the so-called instrumental reflexes, where reinforcement depends on the subject's response to the conditioned signal). Studies have shown that in such situations, reinforcement does not serve to satisfy the need, but to obtain desirable or remove unwanted stimuli. The reflective-evaluative function of emotions is associated with an integral assessment of the relevance and strength of the need experienced by the agent, as well as the likelihood of its satisfaction. This function is one of the elements of the agent learning and decision-making processes.

The switching function of emotions is revealed in the process of competition between the motives of the agent's actions. The decision-making process is associated with the selection of a dominant need from a certain set under the influence of emotional assessment. The compensatory (substitution) function of emotions in natural agents is associated with the special position that emotions occupy in the processes of higher nervous activity. As an active state of specialized brain structures, emotions affect the functioning of systems that regulate behavior, sensory processing, and vegetative (autonomous) functions. The purpose of emotions, in this case, is to weaken the requirements for assessing incoming stimuli and to mobilize the body's resources for a possible response. Typical reaction occurs in a new or unusual environment in which Innenwelt is unable to provide the agent with sufficient information to predict and evaluate. Emotions do not carry information about the environment, but in this case they substitute the existing lack of information in order to adapt the agent to the changed conditions of existence.

### 6.3 0-architecture

Biology Inspired Cognitive Architectures are developed based on the statement about the embodiment of a cognitive agent. The agent's embodiment is expressed by the presence of the Umwelt (as a self-centered individual world of existence), Innenwelt (as an individual universal means of explaining the world), Merkwelt (as part of the Umwelt available for perception) and Werkwelt (as part of Umwelt available for action).

The beginning of the postnatal ontogenesis of natural agents is accompanied by the beginning of the process of cognition. The beginning of Innenwelt development coincides with the opening of the channels of perception. For brevity we will call this moment 0-time.

According to the ideas of constructivism, the beginning of the process of cognition should have some factual basis. In the case under consideration, it can be some initial state of the architecture of the agent's control system. Let's call it 0-architecture — it is the architecture of the control system that exists at the moment 0-time. 0-architecture contains functional kernels of core components. The need to develop the cognitive abilities of the agent implies that 0-architecture must contain the source of this development.

Modeling the dynamics of cognitive activity of the agent requires answers to several questions. Among them, the following: what requirements should be formulated for core components implementation in 0-architecture and what determines them; what is the mechanism and what set of seed processes initializes cognitive evolution.

### 6.3.1 Survival

The changes that are created in Innenwelt are dictated by the needs of the agent. The main need for natural agents at the 0-time is the need for survival. The cognitive natural agent starts its evolution in an aggressive world, which tends to destroy the agent's body. The 0-architecture of natural agents is focused on solving this particular problem. The required speed of agent adaptation depends upon the degree of aggressiveness of the surrounding world. This can be illustrated by comparing insects and mammals: the environment for the existence of insects is much more aggressive, so their adaptation is faster. This example illustrates the influence of Umwelt features on the development of intelligence.

Innenwelt within the cognitive cycle can be seen as the coordinating link between Merkwelt and Werkwelt. The high degree of aggressiveness in the world leads to the construction of simple and effective mental models that allow insects to survive. However, this property of Umwelt blocks the possibility of creating more complex models due to their uselessness. This results in the limitation of the level of agent development, which can be implicitly controlled by having access to the formation of Umwelt.

Core components in 0-architecture should implement a set of unconditioned reflexes. Unconditioned reflexes perform a protective function, and a function of maintaining the constancy of the internal environment of the body (homeostasis). The implementation of a set of unconditioned reflexes in the technical agent is a problem whose solution depends on the conditions of its future existence in the world, on the structure of its body and on the functions of the agent. Unconditioned reflexes in cognitive architectures are built on the basis of a direct connection between a perceptual, motor and emotional component.

The implementation of the motor and perceptual components in 0-architecture has the most complete functionality. This may be explained by their dominant function in the implementation of unconditioned reflexes and in cognitive processes at the sensorimotor stage. Unconditioned reflexes in 0-architecture are realized by processes that do not involve either the intellectual or volitional components. They are implemented by processes that directly link the perceptual, motor and emotional components (see Fig. 1). Processes 1→2 provide a motor response aimed at neutralizing the threat to the agent's existence. Processes 1→8 lead to the appearance of a negative emotional reaction. The processes 7→2 can also be present in the behavior of the agent. They lead to the appearance of a signal reaction (screaming, crying, etc.), the purpose of which is to inform about a situation that threatens health or life.

### 6.3.2 Cognitive development

Another need that natural agents begin to experience at 0-time is the need to interpret the results of perception. This need can be explained as the need to restore the balance with the world, which is indicated by the psychology of constructivism.

0-architecture must contain such a combination of interactions between the components of the control system which is sufficient for the intellectual growth of the agent. We hypothesize that the cognition process is initialized at moment 0-time by some built-in algorithm. We assume that at the beginning of sensorimotor stage of agent development this process is provided by unconditioned reflexes implemented in 0-architecture.

From neurophysiology it is known (for example, [49]) that cognitive functions are determined by the initialization of the orienting-research reaction. This term describes the reaction of animals in experiments where new conditions were presented to them during investigations of conditioned reflexes. Faced with the presence of an unknown factor in a well-known situation, the animal must make a decision about behavior in relation to the new phenomenon. It has been proven that the orienting-research response is based on the orientation reflex, which is a complex response that occurs in humans and animals to a new or unexpected stimulus. From the physiological point of view, this reflex is expressed by the aggregate of motor (increased muscle tone), respiratory (increased breathing rate), autonomic (dilated pupils, the appearance of a galvanic skin reaction, etc.) reactions and by changes in the electrical activity of the brain. The orienting reflex occurs when Innenwelt is unable to provide the agent with a satisfactory explanation for the observed phenomenon. Our hypothesis is that the 0-architecture should contain the implementation of the orienting reflex in addition to realizations of reflexes that maximize the agent's survival (this complex of reflexes can be called survival reflexes).

Above, we proposed to consider two levels of functions provided by the control system: functions of the interface level (used when interacting with Umwelt) and functions of the mental level. The functions of the interface layer are performed by sensors and effectors. In the 0-architecture, these functions are implemented using survival reflexes. One can say that at 0-time the control system is completely implemented in terms of implementation of these functions. However, since Innewelt does not contain information about reality, the interface level turns out to be uncontrollable by the level of mental functions. The task of the agent is to gradually establish an information link between Merkwelt and Werkwelt. It does this by filling the Innenwelt, which is impossible without creating two interfaces at the same time: the sensor data processing interface (Merkwelt $\Rightarrow$ Innenwelt) and the effector control command interface (Innenwelt $\Rightarrow$ Werkwelt). As this task is completed, the level of mental functions gradually gains control over the level of interface functions. This ternary task, which includes the agent's sensory apparatus, effector apparatus and Innenwelt, can also be characterized as the task of constructing a set of functional systems [49].

### 7. COGNITIVE EVOLUTION

Intelligence is a means of adaptation of an agent to the conditions of its existence. The conditions of existence are determined by the needs of the organism, which appear as a result of the agent's interaction with the outside world (not only with the Umwelt). The needs of the organism form the goals of the activity. The connection between the goal and the way to achieve it is expressed through the schemes that exist in the Innenwelt. Schemes are formed as a result of the creation of certain functional systems. The process of creating functional systems involves sensory, effector mechanisms and Innenwelt. In this case, two sets of symbols are formed in Innenwelt, which together implement the interface between the Umwelt interface layer and the mental function layer of the architecture. The first set is created as a result of the agent's interaction with Merkwelt: it is a set of perception symbols. The second set is created as a result of the agent's interaction with Werkwelt: this is a set of action symbols. In addition to these two sets of symbols, Innenwelt should include a means of mapping between them, which is necessary for purposeful activity. These means of mapping should reflect the objectives of the activity. So the goal of the action is a parameter that determines how the Innenwelt connects the symbols of the first and second sets within the framework of a given goal.

In order to map between perception symbols and action symbols, Innenwelt must contain a symbolic expression of goals. However, at 0-time it is absent. At the same time, both of the mentioned sets of symbols are absent. The creation of these sets is initialized using a built-in mechanism based on the use of the orientation reflex. This is carried out on the basis of survival reflexes.

The orienting reflex is the means which, when embedded in 0-architecture, serves to fill the Innenwelt. Figure 2 shows an algorithm that describes the interaction of architecture components in the processing of sensory data. This algorithm is an adaptation of the results of the nervous stimulus model development [57, 58].

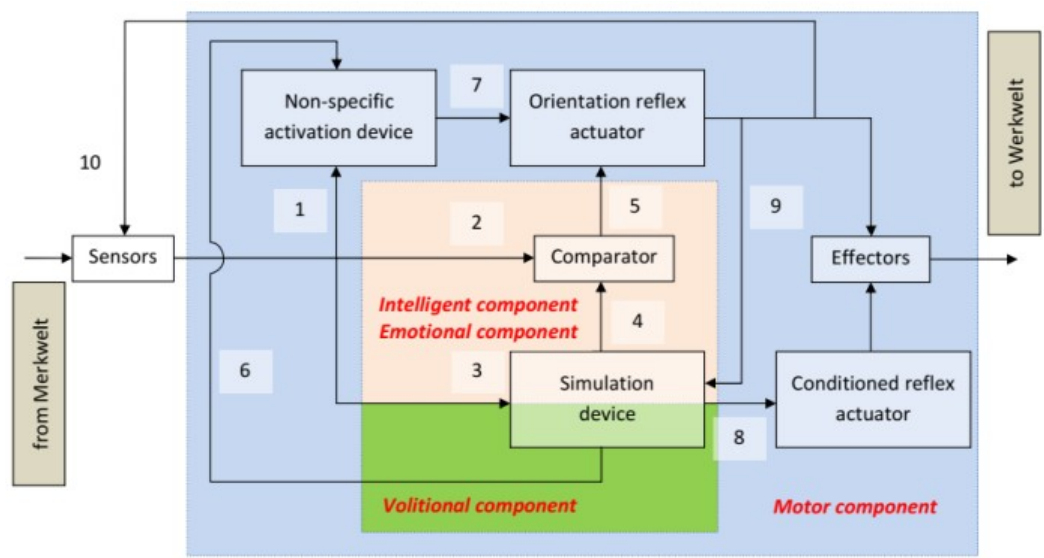

Fig. 2 Interaction of core components during stimulus processing

Merkwelt acts on the sensory component. The signal generated as a result of this action is transferred to three processing units: a simulator (3), a comparator (2), and a nonspecific activation device (1). The non-specific activation device is responsible for the development of an addictive property that leads to a weakening of the activation response. The signal processed by the simulator is sent to a comparator (4), where the signal from the simulator is compared with the signals of the sensory system from the presented stimulus. If they do not match, a mismatch signal is generated and transmitted to the orienting reflex actuator (5). The nervous model of the stimulus is calculated by the simulator. As it is calculated, it enhances its inhibitory effect on the nonspecific activation device at the moment of stimulus action, the model of which has already been created (6). If the deactivation signal (7) is not generated, the operation of the orienting reflex actuator is started.

The appearance of an orientation reflex has an activating effect on the sensory component of the architecture (10), leading to a change in its mode of operation. The activation of the orientation reflex also affects the operation of the simulator (9), initiating an extended and detailed analysis of the parameters and significance of the acting stimulus. Finding that a stimulus is significant leads to the change in Innenwelt. The cognitive architecture creates a new schema in the Innenwelt, or associates the stimulus with existing schemas. The orientation reflex is indirectly involved in the process of creating conditioned reflexes: the model of an indifferent stimulus is connected with the actuator of the conditioned reflex using a simulator (8).

Note that there are two ways to trigger the orienting reflex. In the above explanation, the simulator is able to generate a signal that describes the model of the acting stimulus. The second way of activation is associated with a situation characterized by the impossibility of generating such a signal. It is this situation that describes the operation of 0-architecture at the initial stage of the postnatal period in the development of natural agents. In this case, signals 4, 5, and 6 are absent, and the nonspecific activation device generates signal 7, which initializes the actuator of the orienting reflex. Activation of this device, in turn, activates functions related to the creation of the Innenwelt primary schemes.

The cognitive evolution in the proposed architecture is developing on the basis of schemas embedded in the 0-architecture. They realize unconditioned reflexes. Efficiency is a fundamental requirement for the implementation of these schemes. For this reason, they are all based on the motor component of the architecture. These primary schemes are modified by the orienting reflex. As a result of this modification the first generative schemes of architecture components appear at the initial stage of ontogenesis. Their combinations fill the Innenvelt. The built-in schemes of the motor component are the foundation for building a system of knowledge of an autonomous agent. For natural agents, their presence is the result of the evolutionary process of a definite biological species. A similar evolution can be considered for technical agents. The process of creating embedded schemas can be explored by simulating the co-evolution of agents of several different types within the same technical ecosystem. Consideration of such a problem would make it possible to speak about the solution of SGP and the possibility of using this solution to model the cognitive development of autonomous agents

## 8. SIMULATION OF THE INITIAL STAGE OF POSTNATAL ONTOGENESIS

Cognitive activity is determined by the fact that living beings are able to detect the regularities. Regularities are characterized by the repeatability of the observed phenomena. In order for cognitive activity to be possible, some periodicity in the results of observations is necessary. This thesis concerns the processing of measurement results. In psychology, several stages of sensory information processing are traditionally distinguished. In particular, they include: physical feeling, perception, representation, imagination, and thinking.

Physical feeling is the first step in sensory data processing. It is a reflection of individual properties of reality, acting at the moment on the sensors of a cognitive agent. Physical feeling is a process in which a signal is generated as a result of the impact of the environment on the sensor with the subsequent triggering of a hardware-implemented response to this impact. Hardware implementation can be understood here as a set of physicochemical reactions, or the implementation of algorithms for processing electrical signals, or other mechanisms that can characterize the architecture of a specific sensory system. In biological cognitive systems, a cumulative way of realizing sensations is typical. For a sensation to occur, exposure to a stimulus of a certain intensity is required for a period of time that depends on the intensity.

At the next stage in the processing of sensory data, the resulting sensations must be transformed into information about the surrounding world. The stage of perception is associated with the detailing of the data coming from the sensory component to the level of patterns that characterize Umwelt. At the stage of perception, the processing of sensory data is carried out using the resources of the intelligent component. The neural network of the intelligent component that implements Innenwelt stores the results of previous experience of cognitive activity. At this stage it is included in the processing of data. The formation of a mental representation of a certain object begins with the process of discrimination and then goes through recognition until the complete perception of this object.

The process of discrimination is associated with the use of a certain set of rules for processing sensory data. The use of these rules allows forming hierarchical representations of the Merkwelt.

## 9. DISCUSSION

Existing cognitive architectures have drawbacks that can be eliminated using the proposed architecture. Symbolic architectures use an approach based on the Schema Mechanism: in [51], a formal mathematical apparatus was proposed that implements Piaget's ideas. The main disadvantage of this approach is that developers have to separately search for a solution of the Symbol Grounding Problem in order to be able to use their designs of cognitive architectures. Schema Mechanism does not describe the processes that underlie the learning of autonomous agent. In this regard, developers have to specify not only the components of the cognitive architecture (Attention, Procedural Memory, etc.), but also the rules for their interaction (Planning, Deciding, Scheduling, etc.).

Sub-symbolic architectures allow one to find SGP solution, but have difficulties in implementing high-mental functions. Our approach is to use a functional core to simulate the development of cognitive functions of autonomous agents. The distinctive features of the cognitive architecture proposed in this paper are as follows.

First, the initial set of schemes may be formulated on the basis of a set of reflexes built into the cognitive architecture. However, we do not introduce a symbolic definition for Merkwelt and Werkwelt: perceptual symbols and action symbols are determined in the course of cognitive activity. This means that the Symbol Grounding Problem must be solved during the process of Innenwelt creation.

Second, according to the constructivist approach, at the moment 0-time, the schemes associated with the level of mental functions are absent in our approach. These schemes arise as a result of the cognitive evolution of the initial set of schemes that are associated with the motor component.

Third, we introduce schemas that are directly related to cognitive activity. They are introduced using the orientation reflex. These schemes, being themselves subject to evolution, also participate in the creation of schemes used in the implementation of functions of the mental level.

The most important goal in the field of AGI is the development of control systems for cognitive agents, which, in terms of their intellectual performance, are not inferior, and perhaps even surpass humans. Developmental psychology studies [44, 59] show that the most significant changes in Innenwelt occur at the sensorimotor stage. It is the first one in the postnatal ontogenesis. This stage is characterized by the beginning of the formation of a set of symbols of perception and symbols of action. According to Piaget, this period of development is one of the most important in the creation of human mental abilities. The proposed architecture makes it possible to simulate the process of evolution of cognitive abilities, including the stage of sensorimotor development of autonomous agents. This problem is solved on the basis of the set of unconditioned reflexes, implemented in the 0-architecture. We consider the proposed approach as a development of Drescher's ideas. It allows one to use the results of research in the field of the psychology of early child development and proposals for the creation of appropriate numerical models for simulating the cognitive activity of autonomous agents throughout their life cycle.

The activity of an agent can be phenomenologically characterized as purposeful from the moment when the following conditions are met. The spaces of symbols of perception and symbols of action are not empty. The agent realizes that it is capable of manipulating the action symbols. A mapping from the agent's needs space to the action symbol space has been created. Innenwelt provides a forward and backward mapping between perceptual and action symbol spaces. As a result of these conditions being met, the agent becomes able to formulate: (1) the goals of its actions in terms of symbols of perception and (2) scenarios of its behavior in terms of symbols of action.

Piaget expressed confidence that the complication of the methods used by people in their activities, and the complication of the ways of representing (reflecting) reality, are two fundamentally inseparable aspects of the process of cognitive development. We believe that attempts to directly create cognitive architectures with abilities that are not inferior to human ones should be replaced by the process of modeling the development of cognitive abilities of autonomous agents. The transition to modeling the process of cognition (starting from its earliest stages) makes it possible to study the regularities of this process and use them to develop autonomous cognitive systems with a predefined set of characteristics. At the same time, the question of causal relationships between the processes that characterize each stage of cognitive development remains important.

The use of modern cognitive architectures is limited to the "mechanical" reproduction of functions that, from the point of view of developers, are essential in cognitive activity. Piaget's research showed that the development of intelligence has certain stages that are natural. The formation of mental functions takes place in several stages. The completion of each stage is characterized by a certain qualitative transition. For this reason, the use of "static" structures in the creation of cognitive architectures has no perspective in the reproduction of high-mental functions.

From our point of view, a detailed study of the dynamics of the development of mental functions, starting from the sensorimotor stage, is the only direction leading to the creation of AGI systems. The cognitive architecture discussed in this paper is quite general. We assume that it can be used to create autonomous agents for a wide range of applications. Restrictions on its use will follow from the features of the implementation of the core components and the rules for their interaction.

## 10. CONCLUSIONS

This article deals with the problem of creating a control system for autonomous cognitive agents. The proposed cognitive architecture includes five core components. The interaction of these components makes it possible to simulate changes in the characteristics of an autonomous agent, starting from the sensorimotor stage of development. The evolution of the agent is provided by the built-in core of the control system: 0-architecture.

The proposed 0-architecture model is based on a set of unconditioned reflexes that ensure the agent's survival in the world of embodiment and the initiation of the self-learning process. The paper shows that the process of cognitive development is initialized automatically due to the orienting reflex. The proposed cognitive architecture cannot be attributed to either symbolic or sub-symbolic architectures, since the Symbol Grounding Problem solution is one of the results of modifying the structure of core components in the process of ontogenesis.

In the process of discussing the problems of cognitive development of autonomous agents, we used the terminology of semiotics. This made it possible to carry out the consideration at a sufficiently high level of abstraction. This approach makes it possible to identify a number of theoretical issues that, in our opinion, may be significant for the Developmental Robotics. One such problem is the analysis of the relationship between Merkwelt and Werkwelt. The separation of the world of perception from the world of actions in the analysis of cognitive processes makes it possible to consider the problem of self-learning as a problem of creating functional systems [49].

We consider the problem of cognitive development as the problem of establishing a connection between the Merkwelt and the Werkwelt through the formation of the Innenwelt. Within the framework of the cognitive cycle, this problem is divided into three interrelated tasks: the creation of perception symbols; creating action symbols; and creating forward and backward transformations between symbols of perception and symbols of action. The agent's cognitive development is initialized through the action of the orienting reflex. At an early stage, a decisive role in postnatal ontogenesis is played by reflexes embodied in the 0-architecture of the control system.

This article proposes a research and development approach that would lead to the creation of next-generation intelligent technical systems. A distinctive feature of these systems is their ability to undergo evolutionary change. Implementing such a methodology allows for the full potential of so-called second-nature systems.

We recognize the complexity of solving this problem. Our understanding of the term "second nature" implies the presence of autonomous intelligent agents interacting with each other. The practical implementation of such a system requires addressing the following set of research objectives.

First, the development of a methodology for constructing subjective reality. As we mentioned in previous works, the autonomy of an intelligent agent presupposes the existence of a certain "operational space" - an Innenwelt. This space will be used by the agent to construct a subjective image of reality, create action scenarios, and refine its ability to interact with the world.

Second, the development of a methodology for interaction between autonomous intelligent agents. It is closely related to the cognitive development of an autonomous agent described above. Devices that comprise the second nature, being created on the basis of a single universal cognitive architecture, will differ in their functionality. This means that two different autonomous agents may differ in each of the components of the Umwelt: the Merkwelt, the Werkwelt, and the Innenwelt. This raises a number of interesting research problems related to the mutual adaptation of agents to each other's perceptions and actions.

Third, the development of a methodology for the evolutionary change of a system of autonomous intelligent agents. In this case, by evolution we mean a unified process that includes the following components: changes in the intellectual capabilities of autonomous agents (intelligence improvement); changes in the technical implementation parameters of each autonomous agent (improvement of the agent's physical embodiment); and changes in the interactions between autonomous agents (self-organization of a group of agents).

Fourth, the development of control methods for a system of autonomous intelligent agents. Control of a system whose elements interact is possible in two modes. It can be implemented explicitly, i.e., through direct influence on specific agents. Control can also be implemented implicitly, i.e., by changing key parameters of agent interaction.

Finally, it is necessary to develop a set of technologies capable of implementing the results of solving the above research problems.